\documentstyle[multicol,epsfig,prl,aps]{revtex}

\newcommand{\be}{\begin{equation}}
\newcommand{\ee}{\end{equation}}
\newcommand{\bea}{\begin{eqnarray}}
\newcommand{\eea}{\end{eqnarray}}

\begin{document}

\title{
\begin{flushright}{\normalsize hep-ph/yymmdd \\}
\end{flushright}
Pair production of light pseudoscalar particles in strong inhomogeneous
fields
by the Schwinger mechanism}
 
\author{J.A. Grifols$^a$, Eduard Mass\'o$^a$, Subhendra Mohanty$^b$ 
and K.V. Shajesh$^b$ \\
~}

\address{
$^a$ Grup de F\'{\i}sica Te\`orica and IFAE ,Universitat Aut\`onoma de Barcelona,\\
Bellaterra, Barcelona,Spain.\\
$^b$Physical Research Laboratory ,
Navrangpura, Ahmedabad 380009, India.
} 
\maketitle

\begin{abstract}
We calculate the rate for pair production of light pseudoscalars 
by strong inhomogeneous and static  electric (magnetic) 
fields. We show that, in the case of axions, the stability of atoms over
cosmic lifetimes is jeopardised unless the Peccei-Quinn symmetry breaking scale
 $f_a$  is larger than ${\cal O} (10^{10}\,GeV)$.
\end{abstract}

\vspace{1mm}
\begin{multicols}{2}
The Schwinger mechanism is a non-perturbative process by which an
infinite
number of zero frequency photons can produce an electron-positron pair.
For
example a constant electric field larger than  ${\cal O}(10^{11} V/cm)$ will decay
into
electron-positron pairs, according to the Schwinger
pair production probability 
formula\cite{sch}. In this paper we
consider the pseudoscalar-two-photon coupling. We find that the
pseudoscalar
pair production probability is given by an expression $w\sim g^2
\exp-({\cal O}(m^2 /g) )$ in terms
of  the pseudoscalar-two-photon coupling $g$ and the pseudoscalar
mass $m$. This allows us to constrain the parameter space of $\{m,g\}$
for any pseudoscalar model, based on the stability of electromagnetic fields.    
 In  axion models \cite{axions,axions2} the
effective axion-two-photon 
coupling 
 arise from
axion couplings to charged fermions in a one loop diagram.
This coupling is inversely proportional to the Peccei-Quinn symmetry
beaking
scale $f_a$, and is independent of the mass of the fermion in the loop.
The
axion mass is  inversely proportional to the square of $f_a$. The
pair production formula therefore can be expressed in terms of the
single
parameter $f_a$.

The generic electromagnetic interaction of a pseudoscalar $\phi$ can be written as
(see Fig.1)
\bea
 {\cal L}_I 
= \frac{1}{4} g \phi
	{F}_{\mu \nu}
	\widetilde{F}^{\mu \nu}
\label{a2p}
\eea
with ${F}_{\mu \nu}=\partial_\mu A_\nu - \partial_\nu A_\mu$ and $\widetilde{F}^{\mu
\nu}=  {1\over2} \epsilon^{\mu\nu\alpha\beta}{F}_{\alpha\beta}$.
\begin{center}
\begin{figure}[h] 
\leavevmode
\epsfxsize=130pt
\epsfbox{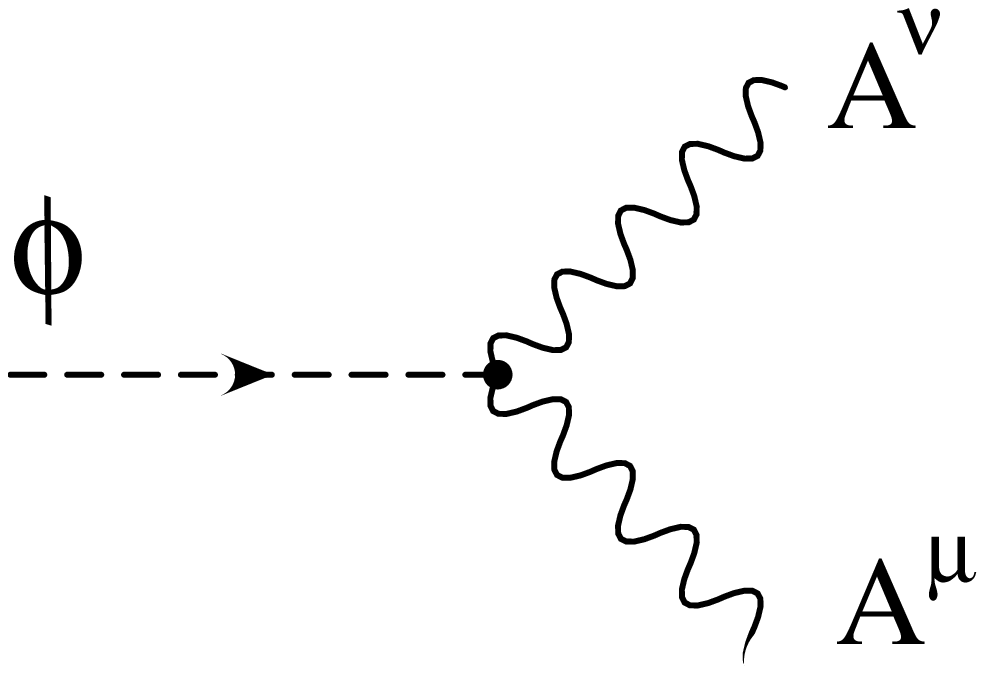}
\label{fig1}
\end{figure}
FIG 1: Axion-two-photon interaction.
\end{center}

We need to evaluate the loop  diagram of the type shown in Fig.2 with
infinite number of zero-frequency photon external legs. The imaginary
part
of this diagram gives the probability for the decay of the external
electromagnetic field.

\begin{center}
\begin{figure}[h]
\leavevmode
\epsfxsize=100pt
\epsfbox{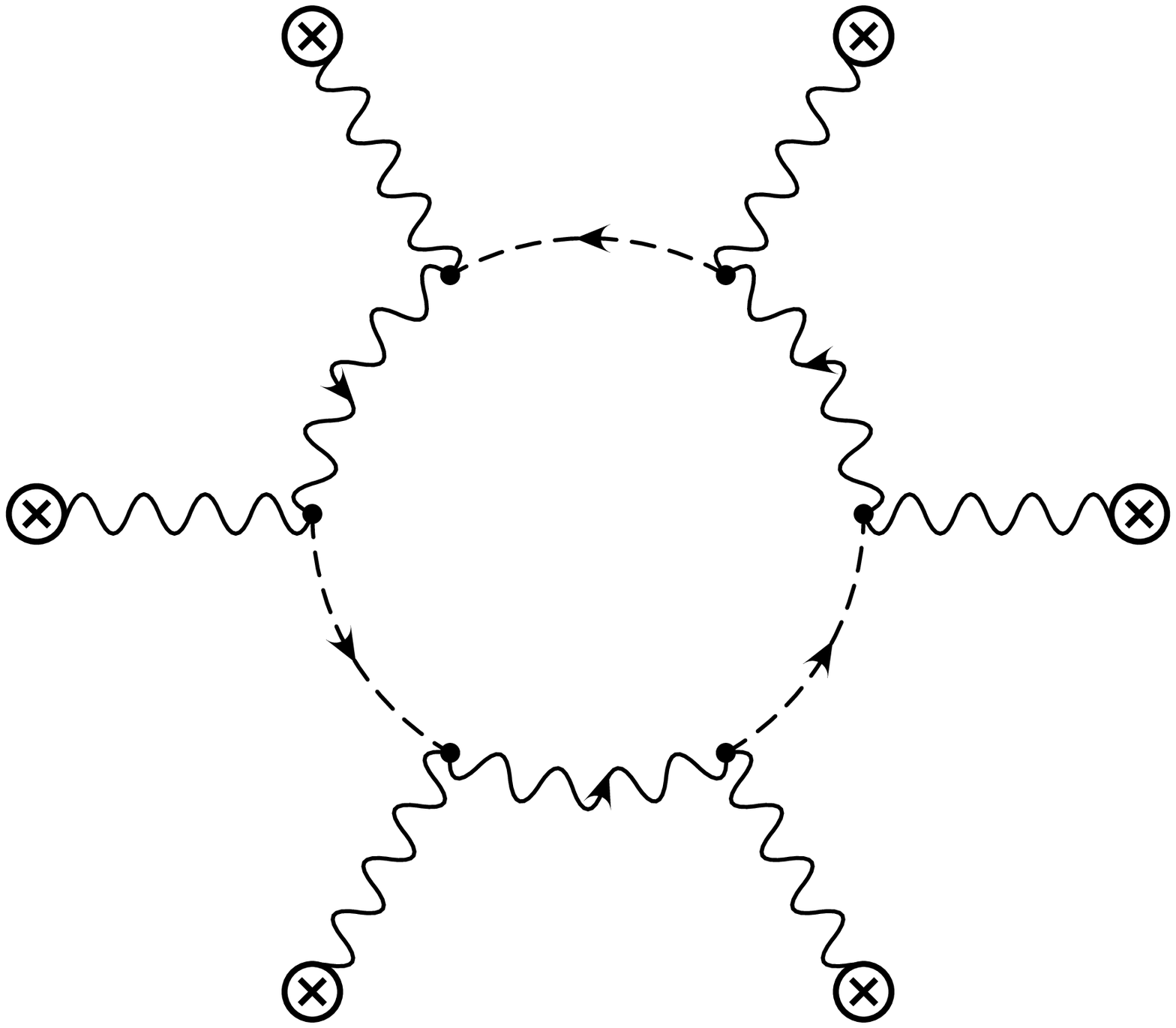}
\label{fig3}
\end{figure}
FIG 2: The effective photon-photon interaction diagram on integrating
out
the axion and a photon field.
\end{center}

We evaluate this diagram in two stages. First
we find the effective vertex for the two axion-two photon vertex which
arises
when we contract the  photon legs  
 between two vertices like (\ref{a2p}). This two axion-two photon
vertex
is of the form
\bea
4 \left( \frac{1}{2} g \phi \right)^2
        k_{\mu}  
        \widetilde{F}^{\mu \nu}
\frac{- i g_{\nu \nu^{\prime}}}{k^2}
        ( - k_{\mu^{\prime}} ) 
        \widetilde{F}^{\mu^{\prime} \nu^{\prime}}
\label{p2a0}
\eea
The factor of $4$ in equation (\ref{p2a0}) is 
for the four  possible ways of joining the photon legs.
Due to the presence of the $k^2$ term in the denominator, (\ref{p2a0})
is non-local.   However, momentum $k$ is integrated over when we calculate the
effective action for the external electromagnetic field.
One can therefore make use of the identity 
\bea
\int d^4 k~ f (k_{\mu} k_{\mu^{\prime}} )
&=& \int d^4 k~
	f \left( \frac{ g_{\mu \mu^{\prime}}}{4} k^2 \right)
\label{eqn4}
\eea
to simplify (\ref{p2a0}) and obtain
\bea
{\cal L^\prime}_{I}
= - \frac{1}{4} g^2 \phi^2
	{F}_{\mu \nu}
	{F}^{\mu \nu} = \frac{1}{2} g^2 \phi^2
	( {\bf{E}}^2 - {\bf{B}}^2 )
\label{lif}
\eea
When the momentum $k$ is integrated over, the effective two axion-two
photon
interaction reduces to a local interaction vertex (Fig 3) given by 
(\ref{lif}).

\begin{center}
\begin{figure}[h]
\leavevmode
\epsfxsize=100pt
\epsfbox{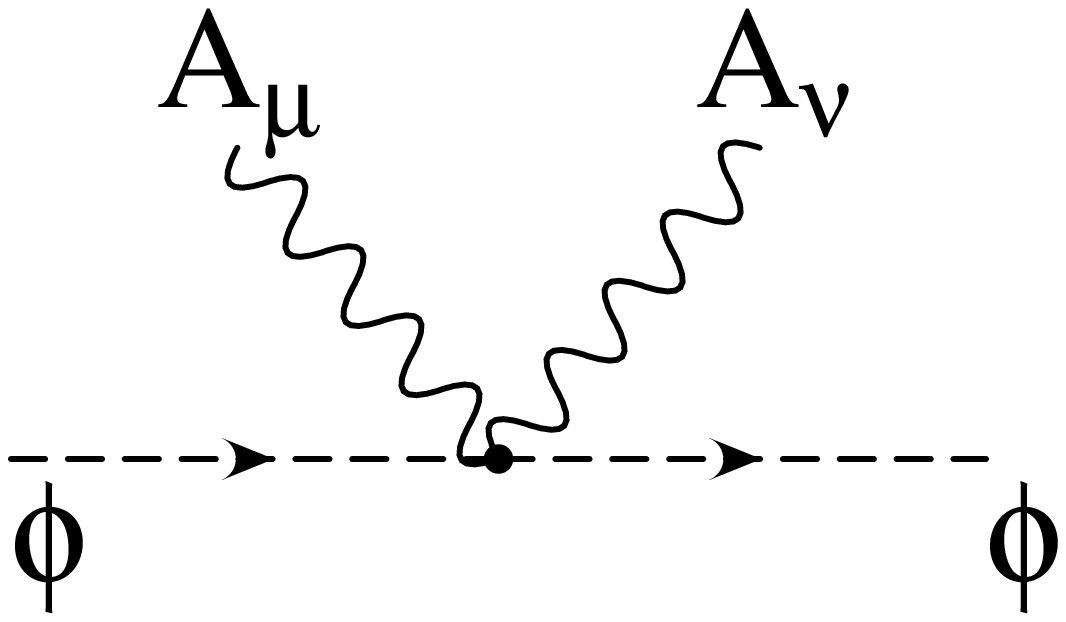}
\label{fig4}
\end{figure}
FIG 3: The effective two axion-two photon interaction in the loop.
\end{center}
Next, to calculate the diagram with infinite insertions of the two-photon
vertices (\ref{lif}) in an  axion loop  we use the
Schwinger method.
Starting with the total action for the axion-photon
system we integrate out the axion fields to determine the effective
action for just the electromagnetic fields
\bea
e^{i S_{eff} [{\bf E},{\bf B} ] }
= \int {\cal D} \phi~
	e^{ i \int d^4 x~ 
	\left\{ \frac{1}{2} \phi 
	( - \partial^2 - m^2 ) \phi
	+ \frac{1}{2} g^2 \phi^2
	( {\bf{E}}^2 - {\bf{B}}^2 )
	\right\} }
\nonumber
\eea
To proceed we shall evaluate ${\cal L}_{eff}$ using a technique
developed in\cite{duff}.
It amounts to expanding the classical background fields about a reference
point, to
compute the (interacting) Green's function in momentum space, to revert
it to
configuration space, and finally obtaining $
{\cal L}_{eff}$ by a parametric
integration. We shall be interested in stationary, inhomogeneous (on the
scale of the pseudoscalar
Compton wavelength) external fields (electric and/or magnetic fields).
The Green's function of the axion field in a background electromagnetic 
field is given by 
\bea
\left[ \partial_{x}^{2} +m^2 - \alpha
	- \beta_i (x-x^{\prime})^i
	- \gamma_{ij}^2 (x-x^{\prime})^{i} (x-x^{\prime})^{j} 
	\right]
&~&~
\nonumber \\ 
\times G(x,x^{\prime})
=
\delta^{4} (x-x^{\prime})
&~&~ 
\eea
\noindent
$(i,j=1,2,3)$ where, as advertised, we have expanded the electromagnetic fields 
to second order about the
reference point $x^{\prime}$. In momentum space this equation reads,
\bea
\left[ - p_0^2 + p_i^2 + m^2 - \alpha 
	+ i \beta_i \frac{\partial}{\partial p_i}
	+ \gamma_{ij}^2 \frac{\partial^2}{\partial p_i \partial p_j}
	\right]
	G(p)
&=& 1
\eea
This equation is solved by making the ansatz
\bea
G(p)
&=& i \int_0^{\infty} ds~ 
	e^{ is (p_0^2 - m^2)}
	e^{ -i {\bf p} \cdot {\bf A} \cdot {\bf p}
	   + {\bf B} \cdot {\bf p} + C }
\eea
and solving for matrix ${\bf A} \equiv A^{ij}$, vector 
${\bf B} \equiv B^i$, and C. The solution is
given by
\bea
\bf{A} 
&=& {1 \over 2}{\mbox{\boldmath $\gamma$}}^{-1} \cdot 
	\tanh 2 \mbox{\boldmath $\gamma$} s
\nonumber \\
\bf{B} 
&=& - { i \over 2} {\mbox{\boldmath $\gamma$}}^{-2} \cdot 
	\left( 1 - {\rm sech}\, 2 \mbox{\boldmath $\gamma$} s \right)
	\cdot \mbox{\boldmath $\beta$}
\nonumber \\
C 
&=& i \alpha s - {1 \over 2} 
	{\rm tr} \ln \cosh 2 \mbox{\boldmath $\gamma$} s
\nonumber \\
&&~~~~~~~~~~~~
	+ {i \over 8}\, \mbox{\boldmath $\beta$} 
	\cdot {\mbox{\boldmath $\gamma$}}^{-3}\cdot
	\left( \tanh 2 \mbox{\boldmath $\gamma$} s 
	- 2 \mbox{\boldmath $\gamma$} s \right)
	\cdot \mbox{\boldmath $\beta$}
\eea
where $tr$ is a trace over indices $i,j$.

Since 
\bea
\frac{\partial {\cal L}_{eff}}{\partial \alpha}
&=& \frac{1}{2i} G(x,x)
\nonumber \\
&=& \frac{1}{2i} \int \frac{d^4 p}{(2\pi)^4} G(p)
\eea
subject to the boundary condition ${\cal L}_{eff}=0$ when ${\bf E}={\bf B}=0$,
we can find ${\cal L}_{eff}$ by gaussian integration to obtain
\end{multicols}
\begin{flushleft} \rule{3.4in}{0.1mm} \end{flushleft}
\vspace{1mm}
\bea
{\cal L}_{eff}
&=& - \frac{1}{32 \pi^2} \int_0^{\infty} \frac{ds}{s^3}~~
	\left[ e^{ -is (m^2 - \alpha)}
	\left( \det \frac{2 {\mbox{\boldmath $\gamma$}} s}{\sinh 2 {\mbox{\boldmath
$\gamma$}} s}
	\right)^{1/2}
	e^{il(s)}
	- e^{-i s m^2}
	\right]
\eea
\vspace{1mm}
\begin{flushright} \rule{3.4in}{0.1mm} \end{flushright}
\begin{multicols}{2}
\noindent
where 
\bea
l(s) = \frac{1}{4}\, \mbox{\boldmath $\beta$}
	\cdot {\mbox{\boldmath $\gamma$}}^{-3} 
	\cdot \left( \tanh \mbox{\boldmath $\gamma$} s
	- \mbox{\boldmath $\gamma$} s \right)
	\cdot \mbox{\boldmath $\beta$} 
\nonumber
\eea
We can now apply this formula to the special case of spherically
symmetric electric fields, like the ones in
atoms. In this case $g^2 E^2$ appearing in the interaction lagrangean is
written
as,
\bea
g^2 E^2 
&=& g^2 E_0^2 + g^2 E_0^{2 \prime} (r-r_0)
	+ \frac{1}{2} g^2 E_0^{2 \prime \prime} (r-r_0)^2 ,
\eea
when expanded about $r_0$. Clearly, $\alpha = g^2 E_0^2$. 
Also, vector $\beta^i$ above reads now
$\mbox{\boldmath $\beta$} = \frac{g^2 E_0^{2 \prime}}{r_0} {\bf r}_0$
and matrix $\mbox{\boldmath $\gamma$}^2$ is, after diagonalization, 
$\mbox{\boldmath $\gamma$}^2 
= {\rm diag}(\gamma_1^2, \gamma_2^2, \gamma_2^2 )$ 
where 
$\gamma_1^2 =\frac{1}{2} g^2 E_0^{2\prime \prime}$
and 
$\gamma_2^2 = \frac{g^2 E_0^{2 \prime}}{2 r_0}$. 
With this input the function $l(s)$ given in equation(12) is
explicitly
\bea
l(s)
&=& r_0^2 \gamma_2^4 \gamma_1^{-3} 
	\left( \tanh \gamma_1 s - \gamma_1 s \right)
\eea
So, finally,
\end{multicols}
\begin{flushleft} \rule{3.4in}{0.1mm} \end{flushleft}
\vspace{1mm}
\bea
{\cal L}_{eff}
&=& - \frac{1}{32 \pi^2} \int_0^{\infty} \frac{ds}{s^3}
	\left[ e^{-is (m^2 - \alpha)}
	\frac{2 \gamma_2 s}{\sinh 2 \gamma_2 s}
	\sqrt{\frac{2 \gamma_1 s}{\sinh 2 \gamma_1 s}}
	\, e^{i l(s)}
	- e^{-i s m^2}
	\right]
\eea
\vspace{1mm}
\begin{flushright} \rule{3.4in}{0.1mm} \end{flushright}
\begin{multicols}{2}
\noindent
Now, when we specialise to the field created by an atomic nucleus
$E^2 =  \frac{Z^2 \alpha}{4 \pi r^4} $, 
we have, 
$\gamma_2^2 = - \frac{g^{2} Z^{2} \alpha}{2 \pi r_0^6}$
and 
$\gamma_1^2 = \frac{5 g^{2} Z^{2}\alpha}{2 \pi r_0^6}$. 
The probability $w$ per unit
volume and unit time for
creation of pseudoscalar pairs is 
$2 Im {\cal L}_{eff}$ and can be obtained from the previous equation
by complex integration. We obtain:
\bea
w &=&
\frac{\widetilde{\gamma_2} \gamma_1}{4 \pi^2} 
	\sum_{n=0}^{\infty} (-1)^n C_{n} \,
	e^{- \kappa (2n+1) \pi}
\label{w}
\eea
Here, 
$\widetilde{\gamma_2}  \equiv  + ( - \gamma_2^2)^{{1}\over{2}}  $, 
and
$\kappa \equiv \frac{1}{2} 
\left( \frac{m^2}{\gamma_1} - \frac{3}{2} \lambda \right)$, 
with
$\lambda \equiv 
\widetilde{\gamma_2}^4 \gamma_1^{-3} r_0^2 $. 
The coefficients $C_n$ in the
asymptotic series above
are given by 
\end{multicols}
\begin{flushleft} \rule{3.4in}{0.1mm} \end{flushleft}
\vspace{1mm}
\bea
C_n 
&& \equiv \int_0^{\pi} du \,
	\frac{e^{-\left( \kappa u + \lambda \cot {u \over 2} \right)}}
	{ \left[ u + (2n+1) \pi \right]^{3\over 2} \,
	\left( \sin u \right)^{1\over 2} \, 
	\sinh \frac{\widetilde{\gamma_2}}{\gamma_1} \,
	\left[ u + (2n+1) \pi \right] }
\eea
\vspace{1mm}
\begin{flushright} \rule{3.4in}{0.1mm} \end{flushright}
\begin{multicols}{2}
\noindent
For the case of axions, the  mass and coupling are related
to the global symmetry breaking scale $f_a$ \cite{axions,axions2}
\bea
g &\simeq& 0.75 \frac{\alpha}{2 \pi f_a}\nonumber\\
m_a &\simeq& 0.6 \, eV \, \frac{10^7 \,GeV}{f_a}
\eea
and one can
estimate the magnitude of the
pair-creation phenomenon in terms of the  single parameter $f_a$.
The asymptotic expansion in the equation for $w$ only makes sense for
$\kappa \ge 1$ and $w$ is
vanishingly small for $\kappa \gg 1$, i.e. for second derivatives of the
fields small on the
scale of $m_a$. Only when fields approach critical values for which
$\kappa \sim {\cal O}(1)$,
pair creation could eventually reach catastrophic rates and produce
vacuum breakdown and
disruption of the external field. Pair creation, which is driven by the
second derivative
of the field, becomes critical (i.e. $\kappa =1$) in a $Z$-atom for
$r_{crit}=({Z f_{a}[GeV]\over 34})^{1/3}\,fm$. Since  $r_{crit}$ must be larger than
a few $fm$, our arguments apply only for  
\bea
f_{a} >  {\cal O} (10^{2}\,GeV) .
\label{lowerl}
\eea

 Let us now constrain $f_{a}$ by realising that the observed cosmic 
abundance of atomic helium agrees with nucleosynthesis up to a few percentage and
therefore we can use the stability over the lifetime
of the Universe of He-atoms.  We 
 consider a thin shell of thickness $\delta r_0 = 0.1r_0$ at critical radius
$r_{crit}$ inside the He-atom. The atom will become ionised due to breakdown of the
electric field in this shell at some time $\tau$ when the
product $Vw\tau \sim 1$ where $V$ is the volume of the shell, and $w$ is as given
in (\ref{w}).

Imposing $\tau > 1.5 \times 10^{10}$ years, and taking into account  (\ref{lowerl}),
we get the following exclusion range for $f_a$:
\bea
10^{2}\,GeV<f_{a}<0.6\times 10^{10}\,GeV
\eea

We note that, for the whole of the exclusion region, the critical radius $r_{crit}\ll
1\AA$, i.e. inside the atom, and the Compton wavelength of the
axion exceeds $10^{3}\,fm$, much larger than the distances over which the field is
inhomogeneous.

We can improve the constraint on $f_a$ by recognising that there are many heavy
atoms which have remained un-ionised over the lifetime of the universe.
 We know from astronomical observations that there are many stars
with lifetime larger than $\tau \sim 10^{10}$ years. This is determined by
comparing the abundance of heavy elements ratio (Th/Eu) in the stars with
that in the solar system. Using the fact that the lifetime of thorium is
$14 \times 10^9$ years it has been estimated that there are stars which
have ages in excess of $12 \times 10^9 $ years \cite{peebles}.  
Since these atomic abundances in the stars are measured from the
intensity of their spectral lines we know that they have not been ionised
due to axion-pair production over times larger than $\tau \sim 10^{10}$
years. However, because $r_{crit}$
grows with $Z$, we cannot use a too large a $Z$ for then $r_{crit}$ would penetrate the atomic
electron cloud and screening would distort the field. We estimate that we can go up to
$Z\sim 20$ and then the bound can be pushed up only to somewhat above
$10^{10}\,GeV$. Anyway, we are not aiming at a very elaborate limit here
\cite{gmm}. We just wish
to point out that to avoid the breakdown of electric fields in atoms by catastrophic
axion pair-creation requires
$f_{a}$ to lie about or beyond the $10^{10}\,GeV$ scale, which is in the ballpark or
already closing the presently still allowed axion-photon coupling window. 

{\it Acknowledgments.} Work partially supported by the CICYT Research Projects
AEN-98-1093 and AEN-98-1116.

\end{multicols}
\vspace{1mm}

\end{document}